\documentstyle[12pt,a4]{article}
\input epsf

\global\arraycolsep=2pt

\begin{document}

\begin{titlepage}

\begin{flushright}
CERN-TH/96-208\\
hep-ph/9608211
\end{flushright}

\vspace{0.5cm}
\begin{center}
\Large\bf QCD Sum-Rule Calculation of the\\
Kinetic Energy and Chromo-Interaction of\\
Heavy Quarks Inside Mesons
\end{center}

\vspace{1.5cm}
\begin{center}
Matthias Neubert\\
{\sl Theory Division, CERN, CH-1211 Geneva 23, Switzerland}
\end{center}

\vspace{1.0cm}
\begin{abstract}
We present a QCD sum-rule determination of the heavy-quark kinetic
energy inside a heavy meson, $-\lambda_1/2 m_Q$, which is consistent
with the field-theory analog of the virial theorem. We obtain
$-\lambda_1\approx (0.10\pm 0.05)~\mbox{GeV}^2$, significantly
smaller than a previous sum-rule result, but in good agreement with
recent determinations from the analysis of inclusive decays. We also
present a new determination of the chromo-magnetic interaction,
yielding $\lambda_2(m_b)=(0.15\pm 0.03)~\mbox{GeV}^2$. This implies
$m_{B^*}^2-m_B^2=(0.60\pm 0.12)~\mbox{GeV}^2$, in good agreement with
experiment. As a by-product of our analysis, we derive the QCD sum
rules for the three form factors describing the meson matrix element
of a velocity-changing current operator containing the gluon
field-strength tensor.
\end{abstract}

\vspace{1.0cm}
\centerline{(Submitted to Physics Letters B)}

\vspace{2.5cm}
\noindent
CERN-TH/96-208\\
August 1996

\end{titlepage}

\section{Introduction}

The physics of hadrons containing a heavy quark simplifies greatly in
the limit where the heavy-quark mass $m_Q$ is taken to infinity. Then
new symmetries of the strong interactions arise, which relate the
long-distance properties of many observables to a small number of
reduced hadronic matrix elements \cite{Shu1}--\cite{review}. A
systematic expansion around the heavy-quark limit has been applied
successfully to learn about the properties of heavy mesons and
baryons, such as their spectroscopy and decays.

A convenient tool to study the implications of the heavy-quark limit
and to perform the $1/m_Q$ expansion is provided by the Heavy-Quark
Effective Theory (HQET) \cite{Geor}, which is constructed by
introducing a velocity-dependent field $h_v(x)$ related to the
original heavy-quark field $Q(x)$ by
\begin{equation}
   h_v(x) = e^{i m_Q v\cdot x}\,\frac{1+\rlap/v}{2}\,Q(x) \,,
\label{redef}
\end{equation}
so that $\rlap/v\,h_v=h_v$. Here $v$ is the 4-velocity of the hadron
containing the heavy quark. The phase redefinition in (\ref{redef})
removes the large ``mechanical'' part $m_Q v$ of the heavy-quark
momentum, which is due to the motion of the heavy hadron. The field
$h_v(x)$ carries the ``residual momentum'' $k=p_Q-m_Q v$, which
arises from the predominantly soft interactions of the heavy quark
with gluons. The effective Lagrangian of the HQET is
\cite{EiHi}--\cite{MRR}
\begin{equation}
   {\cal L}_{\rm eff} = \bar h_v\,i v\!\cdot\!D\,h_v
   + \frac{1}{2 m_Q}\,\bar h_v\,(i D_\perp)^2 h_v
   + C_{\rm mag}(m_Q/\mu)\,\frac{g_s}{4 m_Q}\,
   \bar h_v\,\sigma_{\mu\nu} G^{\mu\nu} h_v + O(1/m_Q^2) \,,
\label{Leff}
\end{equation}
where $D^\mu=\partial^\mu-i g_s A^\mu$ is the gauge-covariant
derivative, and $D_\perp^\mu=D^\mu-(v\cdot D)\,v^\mu$ contains its
components orthogonal to the velocity. The gluon field-strength
tensor is defined as $[D^\mu,D^\nu]=-i g_s G^{\mu\nu}$. The origin of
the operators arising at order $1/m_Q$ in (\ref{Leff}) is most
transparent in the rest frame of the heavy hadron: the first operator
corresponds to the kinetic energy resulting from the residual motion
of the heavy quark inside the hadron (note that $(i D_\perp)^2 =
-(i{\bf D})^2$ in the rest frame), whereas the second operator
describes the magnetic interaction of the heavy-quark spin with the
gluon field. The Wilson coefficient $C_{\rm mag}$ results from
short-distance effects and depends logarithmically on the heavy-quark
mass and on the subtraction scale $\mu$, at which the chromo-magnetic
operator is renormalized \cite{FGL}. As a consequence of a
reparametrization invariance of the HQET, the kinetic operator is not
renormalized \cite{LuMa}.

In many phenomenological applications of the HQET, the forward matrix
elements of the dimension-5 operators in (\ref{Leff}) play a most
significant role. They appear, for instance, in the spectroscopy of
heavy hadrons \cite{Agli}--\cite{Tosh}, in the description of
inclusive decay rates and spectra \cite{Bigi}--\cite{FLS}, as well
as in the normalization of transition form factors at zero recoil
\cite{FaNe}. For the ground-state pseudoscalar and vector mesons,
$M=P$ and $V$, one defines two hadronic parameters, $\lambda_1$ and
$\lambda_2(\mu)$, by\footnote{Another common notation is to define
$\mu_\pi^2=-\lambda_1$ and $\mu_G^2=3\lambda_2$.}
\begin{eqnarray}
   \langle M(v)|\,\bar h_v\,(i D_\perp)^2 h_v\,|M(v)\rangle
   &=& \lambda_1 \,, \nonumber\\
   \langle M(v)|\,\bar h_v\,g_s \sigma_{\mu\nu} G^{\mu\nu} h_v\,
   |M(v)\rangle &=& 2 d_M\lambda_2(\mu) \,,
\label{lamdef}
\end{eqnarray}
where we use a mass-independent normalization of states such that
$\langle M(v)|\,\bar h_v h_v\,|M(v)\rangle=1$. The coefficient $d_M$
takes the values $d_P=3$ and $d_V=-1$ for pseudoscalar and vector
mesons, respectively. The $\mu$ dependence of the parameter
$\lambda_2$ cancels against the $\mu$ dependence of the coefficient
$C_{\rm mag}$ in (\ref{Leff}). The product $C_{\rm mag}\lambda_2$ is
renormalization-group invariant. This quantity can be extracted from
spectroscopy using the relation \cite{FaNe}
\begin{equation}
   \frac 14\,(m_{B^*}^2 - m_B^2) = C_{\rm mag}(1)\,\lambda_2(m_b)
   + O(\Lambda^3/m_b) = 0.12~\mbox{GeV}^2 \,.
\end{equation}
In leading logarithmic approximation, one finds \cite{FGL}
\begin{equation}
   C_{\rm mag}(m_Q/\mu) = \left(
   \frac{\alpha_s(m_Q)}{\alpha_s(\mu)} \right)^{3/\beta_0} \,,
\end{equation}
where $\beta_0=11-\frac 23 n_f$ is the first coefficient of the
$\beta$ function. Using this result, we obtain
\begin{equation}
   \lambda_2(m_b)\simeq 0.12~\mbox{GeV}^2 \,,\quad
   \lambda_2(\mu_0)\simeq 0.15~\mbox{GeV}^2 \,,
\label{lam2val}
\end{equation}
where we have used the values $\alpha_s(m_b)=0.21$ and
$\alpha_s(\mu_0)=0.4$, corresponding to a low renormalization point
$\mu_0\approx 1~\mbox{GeV}$.

Although spectroscopic relations may be used to extract the
differences of the matrix elements of the kinetic operator between
different hadron states, the value of the parameter $\lambda_1$
itself cannot be determined from spectroscopy. Indeed, at present it
is not even known whether $\lambda_1$ is a ``physical'' parameter in
the sense that it can be defined unambiguously in a non-perturbative
way. It may be that any such definition is intrinsically ambiguous
because of the presence of infrared renormalons; however, it may also
happen that for some fortuitous reason renormalons are absent in the
case of the kinetic operator \cite{BeBr,MNS}. From the practitioner's
point of view, $\lambda_1$ becomes a useful parameter once a scheme
for treating perturbative corrections in the HQET has been specified.
Here, we shall work in the $\overline{\mbox{\sc ms}}$ subtraction
scheme and assume that our value of $\lambda_1$ is used in connection
with theoretical expressions that have one-loop perturbative
accuracy.

Theoretical information about the parameter $\lambda_1$ can be
obtained from a non-perturbative evaluation of the first matrix
element in (\ref{lamdef}). Using QCD sum rules in the HQET, Ball and
Braun have derived the value $-\lambda_1=(0.52\pm 0.12)~\mbox{GeV}^2$
\cite{BaBr}, which has been adopted subsequently in many
phenomenological analyses. This value is surprisingly large; it
implies an average momentum of the heavy quark inside the meson of
order 600--800~MeV. In fact, an earlier QCD sum-rule calculation
using a less sophisticated approach had given the smaller value
$-\lambda_1=(0.18\pm 0.06)~\mbox{GeV}^2$ \cite{ElSh}. A theoretical
argument in favour of a smaller value of the kinetic energy was
presented in Ref.~\cite{virial}: the field-theory analog of the
virial theorem relates the first matrix element in (\ref{lamdef}) to
a matrix element of an operator containing the gluon field-strength
tensor, making explicit an ``intrinsic smallness'' of $\lambda_1$. On
the other hand, a large value of $-\lambda_1$ was argued for by
Voloshin and by Bigi et al., who derived the lower bound
$-\lambda_1>3\lambda_2\simeq 0.36~\mbox{GeV}^2$ using first a
quantum-mechanical reasoning \cite{Volo} and later field-theoretical
arguments based on zero-recoil sum rules \cite{BSUV}. Recently,
however, it was shown that this bound is weakened significantly by
higher-order perturbative corrections \cite{KLWG}. Thus, small values
of $-\lambda_1$ can no longer be excluded a priori. Several authors
have attempted to extract $\lambda_1$ (together with the ``binding
energy'' $\bar\Lambda$) from a combined analysis of inclusive decay
rates and moments of the decay spectra in beauty and charm decays
\cite{LiNi}--\cite{Cher}. The most recent of such analyses give
values $-\lambda_1\approx 0.1$--0.2~GeV$^2$. A value of $\lambda_1$
has also been extracted using a lattice simulation of the HQET,
yielding $-\lambda_1=(-0.09\pm 0.14)~\mbox{GeV}^2$ \cite{GiMS}.

In view of these developments, it seems worth while to reconsider the
problem of calculating $\lambda_1$ and $\lambda_2$ using QCD sum
rules. Here we present the results of a new analysis, which is based
on the virial theorem \cite{FaNe,virial}.\footnote{Preliminary
results of this analysis have been presented in
Ref.~\protect\cite{QCD94}.} Our approach has the advantage that both
parameters are determined simultaneously from the zero-recoil (i.e.\
equal-velocity) limit of the QCD sum rule for the matrix element of a
local dimension-5 operator between two meson states moving at
different velocities. As a result, $\lambda_1$ and $\lambda_2$ are
obtained from a set of two sum rules with very similar systematic
uncertainties.

\section{Derivation of the Sum Rules}

The central object of our study is the meson matrix element of a
local dimension-5 operator containing two heavy-quark fields at
different velocities. Using the covariant tensor formalism of the
HQET \cite{FGGW}, we write
\begin{equation}
   \langle M'(v')|\,\bar h_{v'}\Gamma\,i g_s G^{\mu\nu} h_v\,
   |M(v)\rangle = - \mbox{Tr}\left\{ \phi^{\mu\nu}(v,v')\,
   \overline{\cal M}'(v') \Gamma {\cal M}(v) \right\} \,,
\label{phidef}
\end{equation}
where $\Gamma$ is an arbitrary Dirac matrix, and
\begin{equation}
   {\cal M}(v) = \frac{1+\rlap/v}{2\sqrt 2}\,\cases{
   \gamma_5 \,; &pseudoscalar meson $P(v)$ \cr
   \rlap/\epsilon \,; &vector meson $V(v,\epsilon)$ \cr}
\end{equation}
is a matrix representing the spin wave-function of a ground-state
meson $M$ moving at velocity $v$. The most general decomposition of
the tensor form factor $\phi^{\mu\nu}$ consistent with Lorentz
covariance and heavy-quark symmetry reads
\begin{equation}
   \phi^{\mu\nu}(v,v') = \phi_1(w)\,(v^\mu v'^\nu - v^\nu v'^\mu)
   + \phi_2(w) \left[ (v-v')^\mu\gamma^\nu
   - (v-v')^\nu\gamma^\mu \right] + \phi_3(w)\,i\sigma^{\mu\nu} \,,
\label{phidec}
\end{equation}
where $w=v\cdot v'$. For equal velocities ($w=1$) only the last term
appears, and comparing with (\ref{lamdef}) we obtain the
normalization condition
\begin{equation}
   \phi_3(1) = \lambda_2 \,.
\label{phi3norm}
\end{equation}
The equations of motion of the HQET imply another normalization
condition for a certain combination of the invariant functions
$\phi_i(w)$. It reads \cite{FaNe}
\begin{equation}
   3 \phi_1(1) - 3 \phi_2(1) - \frac 32\,\phi_3(1) = - \lambda_1 \,.
\label{phinorm}
\end{equation}
This relation is remarkable in that it relates the matrix element of
the kinetic operator in (\ref{lamdef}) to a matrix element of the
gluon field-strength tensor, in accordance with the picture that the
residual motion of the heavy quark inside the meson is caused by its
interactions with gluons. Eq.~(\ref{phinorm}) can be interpreted as
the field-theory analog of the virial theorem, which relates the
kinetic energy to a matrix element of the ``electric'' components of
the gluon field \cite{virial}.

We shall now derive the Laplace sum rules for the invariant functions
$\phi_i(w)$. The analysis proceeds in complete analogy to that of the
Isgur--Wise function. For a detailed discussion of the procedure and
notations, the reader is referred to Ref.~\cite{IWfun}. We consider,
in the HQET, the 3-point correlation function of the local operator
appearing in (\ref{phidef}) with two interpolating currents for the
ground-state heavy mesons:
\begin{eqnarray}
   &&\int\mbox{d}x\,\mbox{d}y\,e^{i(k'\cdot y-k\cdot x)}\,
    \langle\,0\,|\,\mbox{T}\left\{
    \bar q\,\overline{\Gamma}_{M'} h_{v'}(y) ,
    \bar h_{v'}\Gamma\,i g_s G^{\mu\nu} h_v(0) ,
    \bar h_v\Gamma_M\,q(x) \right\} |\,0\,\rangle \nonumber\\
   &&\quad = \mbox{Tr}\left\{ \Phi^{\mu\nu}(v,v',k,k')\,
    \overline{\Gamma}_{M'} \frac{1+\rlap/v'}{2}\,\Gamma\,
    \frac{1+\rlap/v}{2}\,\Gamma_M \right\} \,,
\label{correl}
\end{eqnarray}
where $k$ and $k'$ are the external momenta. Depending on the choice
$\Gamma_M=\gamma_5$ or $\Gamma_M=\gamma^\alpha-v^\alpha$, the
heavy-light currents interpolate pseudoscalar or vector mesons,
respectively. The Dirac structure of the correlator, as shown in the
second line, is a consequence of the Feynman rules of the HQET. The
quantity $\Phi^{\mu\nu}$ obeys a decomposition analogous to
(\ref{phidec}), with coefficient functions $\Phi_i(\omega,\omega',w)$
that are analytic in the ``residual energies'' $\omega=2 v\cdot k$
and $\omega'=2 v'\cdot k'$, with discontinuities for positive values
of these variables. These functions also depend on the velocity
transfer $w=v\cdot v'$.

The idea of QCD sum rules is to relate a theoretical approximation to
the Operator Product Expansion (OPE) of the above correlator to a
hadronic representation of the same correlator in terms of physical
intermediate states. The lowest-lying states are the ground-state
mesons $M(v)$ and $M'(v')$ associated with the heavy-light currents.
They lead to a double pole located at $\omega=\omega'= 2\bar\Lambda$,
where $\bar\Lambda=m_M-m_Q$ is the ``effective mass'' of the
ground-state mesons in the HQET \cite{Luke}. The residue of this
double pole is proportional to the invariant functions $\phi_i(w)$.
We find
\begin{equation}
   \Phi_i^{\rm pole}(\omega,\omega',w)
   = \frac{\bar\Lambda\,\phi_i(w)\,F^2}
   {(\omega-2\bar\Lambda+i\epsilon)
    (\omega'-2\bar\Lambda+i\epsilon)} \,,
\label{pole}
\end{equation}
where $F$ is the meson decay constant in the HQET ($F\simeq
f_M\sqrt{m_M}$). In the deep Euclidean region the correlator can be
calculated perturbatively because of asymptotic freedom. The main
assumption behind QCD sum rules is that, at the transition from the
perturbative to the non-perturbative regime, confinement effects can
be described by including the leading power corrections in the OPE.
They are proportional to vacuum expectation values of local
quark--gluon operators, the so-called condensates \cite{SVZ}.
Following the standard procedure, we write the theoretical
expressions for $\Phi_i$ as double dispersion integrals and perform a
Borel transformation in the variables $\omega$ and $\omega'$. This
eliminates possible subtraction polynomials and yields an exponential
damping factor in the dispersion integrals, which suppresses the
contributions from excited states. Because of heavy-quark symmetry,
it is natural to set the associated Borel parameters equal:
$\tau=\tau'\equiv 2 T$. Following Refs.~\cite{IWfun,BlSh}, we then
introduce new variables $\omega_+=\frac 12(\omega+\omega')$ and
$\omega_-=\omega-\omega'$, perform the integral over $\omega_-$, and
employ quark--hadron duality to equate the remaining integral over
$\omega_+$ up to a ``continuum threshold'' $\omega_c$ to the Borel
transform of the double-pole contribution in (\ref{pole}). This
yields the Laplace sum rules
\begin{equation}
   \phi_i(w)\,F^2\,e^{-2\bar\Lambda/T}
   = \int\limits_0^{\omega_c}\!\mbox{d}\omega_+\,e^{-\omega_+/T}\,
   \bar\rho_i(\omega_+,w) \,.
\label{sumrul}
\end{equation}
The spectral densities $\bar\rho_i(\omega_+,w)$ arise after
integration of the double spectral densities over $\omega_-$.

\begin{figure}
\epsfxsize=10cm
\centerline{\epsffile{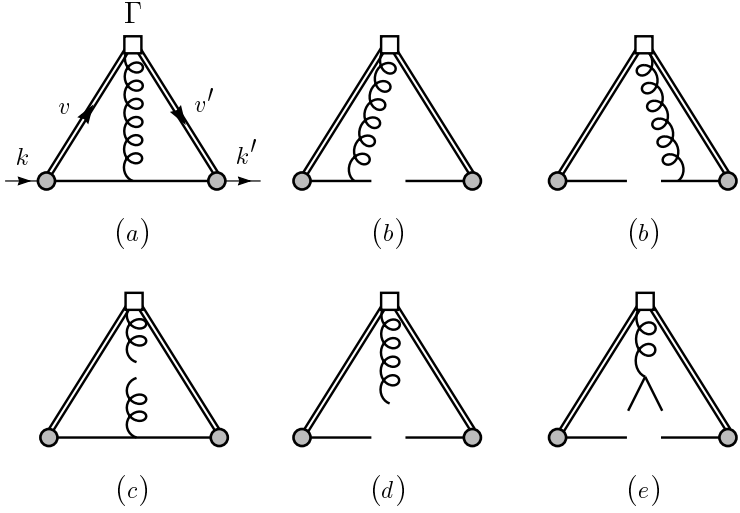}}
\caption{Non-vanishing diagrams for the 3-point correlator: (a)
perturbative contribution, (b) quark-condensate, (c)
gluon-condensate, (d) mixed-condensate, and (e) four-quark condensate
contributions. The velocity-changing current operator is denoted by a
white square, the interpolating meson currents by gray circles.
Heavy-quark propagators are drawn as double lines.}
\label{fig:diagrams}
\end{figure}

As pointed out above, the theoretical expressions on the right-hand
side of the sum rules consist of perturbative and non-perturbative
contributions. The leading terms in the OPE arise from the diagrams
shown in Fig.~\ref{fig:diagrams}. In our analysis, we shall include
the non-perturbative contributions of the quark condensate
$\langle\bar q q\rangle$, the gluon condensate $\langle\alpha_s
G^2\rangle$, and the mixed quark--gluon condensate $\langle\bar
q\,g_s\sigma_{\mu\nu} G^{\mu\nu} q\rangle\equiv m_0^2\,\langle\bar q
q\rangle$. For a consistent calculation at order $\alpha_s$, we
calculate the Wilson coefficients of the quark and gluon condensates
to one-loop order, and the coefficient of the mixed condensate at
tree level. At higher orders in the OPE, one encounters a
proliferation of condensates whose values are essentially unknown.
The terms of dimension six, in particular, consist of four-quark and
three-gluon condensates. For an estimate of such contributions we
include the effects of four-quark condensates, which arise from the
diagram shown in Fig.~\ref{fig:diagrams}(e). The calculation of the
condensate terms is most conveniently performed using the
fixed-point gauge $x\cdot A(x)=0$ with the origin chosen at the
position of the velocity-changing heavy-quark current. The most
complicated part of the calculation is, however, to evaluate the
perturbative contribution of the two-loop diagram shown in
Fig.~\ref{fig:diagrams}(a). We have calculated this diagram using the
techniques developed in Ref.~\cite{2loop}. Our results for the
Laplace sum rules are:
\begin{eqnarray}
   \phi_1(w)\,F^2\,e^{-2\bar\Lambda/T}
   &=& \frac{2\alpha_s T^5}{\pi^3}\left( \frac{2}{w+1} \right)^2
    \delta_4(\omega_c/T) \,, \nonumber\\
   \phi_2(w)\,F^2\,e^{-2\bar\Lambda/T}
   &=& -\frac{2\alpha_s T^5}{\pi^3}\,\frac{2}{w+1}\,
    \delta_4(\omega_c/T) + \frac{4\alpha_s T^2}{3\pi}\,
    \langle\bar q q\rangle\,\delta_1(\omega_c/T) \nonumber\\
   &&\mbox{}- \frac{T}{48\pi}\,\frac{2}{w+1}\,
    \langle\alpha_s G^2\rangle\,\delta_0(\omega_c/T)
    + \frac{\langle O_6\rangle}{12 T} \,, \nonumber\\
   \phi_3(w)\,F^2\,e^{-2\bar\Lambda/T}
   &=& \frac{4\alpha_s T^5}{\pi^3}\,\frac{2}{w+1}\,
    \delta_4(\omega_c/T) - \frac{8\alpha_s T^2}{3\pi}\,
    \langle\bar q q\rangle\,\delta_1(\omega_c/T) \nonumber\\
   &&\mbox{}+ \frac{T}{24\pi}\,\frac{2}{w+1}\,
    \langle\alpha_s G^2\rangle\,\delta_0(\omega_c/T)
    - \frac{m_0^2\langle\bar q q\rangle}{12}
    - \frac{\langle O_6\rangle}{12 T} \,.
\label{phisr}
\end{eqnarray}
The functions $\delta_n(\omega_c/T)$ arise from the continuum
subtraction and are given by
\begin{equation}
   \delta_n(x) = \frac{1}{n!}\int\limits_0^x\!\mbox{d}t\,
   t^n e^{-t} = 1 - e^{-x} \sum_{k=0}^n \frac{x^k}{k!} \,.
\end{equation}
The four-quark condensate $\langle Q_6\rangle$ is defined as
\begin{equation}
   \langle Q_6\rangle = g_s^2\,\langle\bar q\gamma^\mu t_a q\,
   \sum_f \bar f\gamma_\mu t_a f\rangle
   \equiv - \frac{16\pi}{9}\,\kappa\,\alpha_s
   \langle\bar q q\rangle^2 \,.
\end{equation}
Assuming factorization of the four-quark operator \cite{SVZ}
corresponds to setting $\kappa=1$.

In the next step, we evaluate the sum rules in (\ref{phisr}) for
$w=1$ and use the normalization conditions (\ref{phi3norm}) and
(\ref{phinorm}) to obtain the Laplace sum rules for the hadronic
parameters $\lambda_1$ and $\lambda_2$. This leads to
\begin{eqnarray}
   -\lambda_1\,F^2\,e^{-2\bar\Lambda/T}
   &=& \frac{6\alpha_s T^5}{\pi^3}\,\delta_4(\omega_c/T)
    + \frac{m_0^2\langle\bar q q\rangle}{8}\,(1-\varepsilon_6)
    \,, \nonumber\\
   \lambda_2\,F^2\,e^{-2\bar\Lambda/T}
   &=& \frac{4\alpha_s T^5}{\pi^3}\,\delta_4(\omega_c/T)
    - \frac{8\alpha_s T^2}{3\pi}\,\langle\bar q q\rangle\,
    \delta_1(\omega_c/T) \nonumber\\
   &&\mbox{}+ \frac{T}{24\pi}\,\langle\alpha_s G^2\rangle\,
    \delta_0(\omega_c/T) - \frac{m_0^2\langle\bar q q\rangle}{12}\,
    (1+\varepsilon_6) \,,
\label{lamsr}
\end{eqnarray}
where
\begin{equation}
   \varepsilon_6 = - \frac{16\pi}{9}\,\kappa\,
   \frac{\alpha_s\langle\bar q q\rangle}{m_0^2\,T} \,.
\end{equation}
For all reasonable values of the parameters, $\varepsilon_6$ is of
order a few per cent, which is much less than the uncertainty in the
parameter $m_0^2$. Therefore, the contribution of the four-quark
condensate can be safely neglected in the numerical analysis, and we
shall set $\varepsilon_6=0$ hereafter. The sum rule for $\lambda_2$
(without the contribution of the four-quark condensate) coincides
with the result derived by Ball and Braun \cite{BaBr}; our sum rule
for $\lambda_1$ is new. Notice that the sum rule for $\lambda_1$ does
not receive contributions from the quark and gluon condensates. This
is a consequence of the fact that (in the fixed-point gauge) the
light quark interacts only with the magnetic components of the gluon
field \cite{Novi}, whereas the combination of form factors defining
$\lambda_1$ in (\ref{phinorm}) corresponds to a matrix element of the
electric components \cite{virial}.

For the evaluation of the sum rules, it is convenient to eliminate
the explicit dependence on the parameters $F$ and $\bar\Lambda$ by
using the well-known sum rule for the correlator of two heavy-light
currents \cite{Shur,IWfun,BBBD}:\footnote{Since the leading terms in
(\protect\ref{phisr}) are proportional to $\alpha_s$ and we have no
control over the $O(\alpha_s^2)$ corrections, we do not include the
known $O(\alpha_s)$ corrections in (\protect\ref{Fsr}) for
consistency.}
\begin{equation}
   F^2\,e^{-2\bar\Lambda/T} = \frac{3 T^3}{4\pi^2}\,
   \delta_2(\omega_c/T) - \langle\bar q q\rangle
   + \frac{m_0^2\langle\bar q q\rangle}{4 T^2} \,.
\label{Fsr}
\end{equation}
Dividing the sum rules in (\ref{lamsr}) by the sum rule in
(\ref{Fsr}), we obtain expressions for the parameters $\lambda_1$ and
$\lambda_2$ as functions of the Borel parameter $T$ and the continuum
threshold $\omega_c$. This procedure reduces the systematic
uncertainties in the calculation. Moreover, it eliminates the
dependence on the parameter $\bar\Lambda$, which is known to suffer
from a renormalon ambiguity problem \cite{BeBr,BiSh}. For the QCD
parameters entering the theoretical expressions, we take the standard
values \cite{SVZ}
\begin{eqnarray}
   \langle\bar q q\rangle &=& -(0.23\pm 0.02)^3~\mbox{GeV}^3
    \,, \nonumber\\
   \langle\alpha_s GG\rangle &=& (0.04\pm 0.02)~\mbox{GeV}^4 \,,
    \nonumber\\
   m_0^2 &=& (0.8\pm 0.2)~\mbox{GeV}^2 \,,
\label{cond}
\end{eqnarray}
as well as $\alpha_s=0.4$. These values refer to a renormalization
scale $\mu_0\simeq 2\bar\Lambda \approx 1$ GeV, which is appropriate
for evaluating QCD sum rules in the HQET. We shall comment below on
the sensitivity of our result to the choice of the condensate
parameters.

The sum-rule parameters $\omega_c$ and $T$ should, in principle, be
determined in a self-consistent way by requiring optimal stability of
the results under variations of the Borel parameter inside the region
where the theoretical calculations are reliable. For too small values
of $T$, the OPE diverges, whereas for large values of $T$ the
contributions to the sum rules from higher resonance states become
more and more important. Unfortunately, the continuum-contamination
problem is rather severe in the case of sum rules for the matrix
elements of higher-dimensional operators such as $\lambda_1$ and
$\lambda_2$. This is exemplified by the leading perturbative
contribution to the correlator in (\ref{correl}), which is
proportional to
\begin{equation}
   \frac{1}{4!\,T^5} \int\limits_0^\infty\!\mbox{d}\omega_+\,
   \omega_+^4\,e^{-\omega_+/T} = \delta_4(\omega_c/T)
   + \Big[ 1 - \delta_4(\omega_c/T) \Big] \,.
\end{equation}
The first term on the right-hand side is assigned to the
ground-state, whereas the second term is removed in the continuum
subtraction. For the central values $T=0.9$~GeV and
$\omega_c=2.0$~GeV determined below, the ground-state contribution is
only 7.5\% of the total perturbative contribution. For comparison, in
the case of the sum rule in (\ref{Fsr}), the leading perturbative
contribution to the correlator is proportional to
$\delta_2(\omega_c/T) + [1-\delta_2(\omega_c/T)]$, and the
ground-state contribution amounts to 38\%. For this reason, it is
better to determine the allowed regions for the parameters $\omega_c$
and $T$ by requiring stability of the sum rule (\ref{Fsr}) for the
meson decay constant, and then to use the same values in the
evaluation of the sum rules for $\lambda_1$ and $\lambda_2$
\cite{BaBr}. One finds that $\omega_c=(2.0\pm 0.3)$~GeV, and the
``stability window'' for the Borel parameter is
$0.6~\mbox{GeV}<T<1.2~\mbox{GeV}$ \cite{IWfun,BBBD}.

\begin{figure}
\epsfxsize=10cm
\centerline{\epsffile{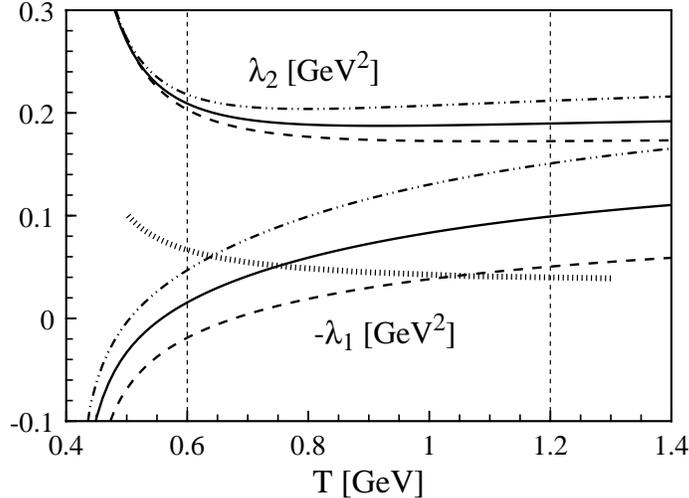}}
\caption{Sum-rule results for the parameters $-\lambda_1$ (lower
curves) and $\lambda_2$ (upper curves). For each quantity, the three
curves correspond to the following values of the continuum threshold:
$\omega_c=1.7$~GeV (dashed), 2.0~GeV (solid), 2.3~GeV (dash-dotted).
The vertical dashed lines show the sum-rule window.}
\label{fig:lam12}
\end{figure}

Using these ranges of parameters, together with the central values of
the condensates given in (\ref{cond}), we obtain the results shown in
Fig.~\ref{fig:lam12}. The sum rule for the parameter describing the
chromo-magnetic interaction of the heavy quark exhibits very good
stability. Taking an average over the sum-rule window, we obtain
\begin{equation}
   \lambda_2(\mu_0) = (0.19\pm 0.02\pm 0.02)~\mbox{GeV}^2 \,,
\label{lam2mu0}
\end{equation}
where the first error reflects the variation with $\omega_c$ and $T$,
while the second error takes into account the uncertainty in the
values of the vacuum condensates. When evolved to a high
renormalization point, our result corresponds to
$\lambda_2(m_b)=(0.15\pm 0.03)~\mbox{GeV}^2$, which is in good
agreement with the value in (\ref{lam2val}) extracted from
spectroscopy. A very similar result has been obtained by Ball and
Braun using the same approach \cite{BaBr}, and by the present author
using a different analysis based on two-point sum rules \cite{subl}.

The stability of the sum rule for the kinetic-energy parameter
$\lambda_1$ is not quite as good. The reason is that the condensate
contribution has the opposite sign of the perturbative contribution.
Inside the allowed parameter space for $\omega_c$ and $T$, we find
values for $-\lambda_1$ ranging from $-0.02~\mbox{GeV}^2$ to
$+0.15~\mbox{GeV}^2$. The dependence on the Borel parameter is
strongest in the region of low $T$ values, where the contribution of
the mixed condensate becomes very large. If we restrict ourselves to
the region of larger $T$ values by requiring that the condensate term
be less than 50\% of the perturbative contribution, we find that the
region below the hatched line in Fig.~\ref{fig:lam12} is excluded.
This leads to
\begin{equation}
   -\lambda_1\approx (0.10\pm 0.05\pm 0.02)~\mbox{GeV}^2 \,,
\label{lam1val}
\end{equation}
where the second error reflects again the dependence on the
condensate parameters. It must be stressed that because of the
relatively poor stability the sum-rule prediction for $\lambda_1$ is
affected by systematic uncertainties that may be underestimated by
the error quoted in (\ref{lam1val}). Keeping this reservation (which
applies equally to previous sum-rule determinations of $\lambda_1$)
in mind, we note that our value in (\ref{lam1val}) implies an average
momentum of the heavy quark inside the meson of order 200--400~MeV,
which appears to us to be a reasonable value. Clearly, our result is
much smaller than the value $-\lambda_1=(0.52\pm 0.12)~\mbox{GeV}^2$
obtained by Ball and Braun \cite{BaBr}; indeed, we find
$-\lambda_1<\lambda_2(\mu_0)$ for all choices of the parameters. We
shall comment below on the difference between their approach and
ours.

\begin{figure}
\epsfxsize=10cm
\centerline{\epsffile{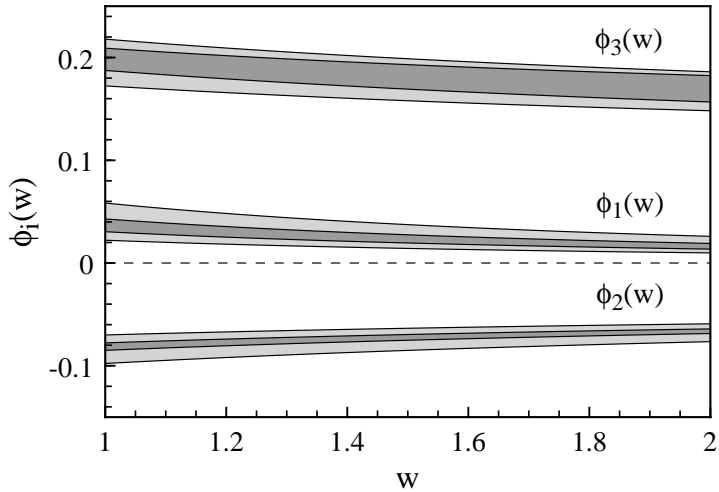}}
\caption{Sum-rule results for the functions $\phi_i(w)$. The width of
the bands reflects the variation of the results with the continuum
threshold ($1.7~\mbox{GeV}<\omega_c<2.3~\mbox{GeV}$ for the light
bands, and $\omega_c=2$~GeV for the inner, dark bands) and the Borel
parameter ($0.6~\mbox{GeV}<T<1.2~\mbox{GeV}$).}
\label{fig:phifuns}
\end{figure}

Although our main focus was to derive sum rules for the parameters
$\lambda_1$ and $\lambda_2$, the invariant form factors $\phi_i(w)$
defined in (\ref{phidef}) may be of some interest as well. For
instance, the combination $f(w)=3\phi_3(w)-2(w-1)\phi_2(w)$ appears
in the analysis of non-factorizable contributions to class-I
non-leptonic two-body decays such as $\bar B^0\to D^+\pi^-$
\cite{Blok}. Therefore, we find it worth while to study the $w$
dependence of these form factors using the sum rules in (\ref{phisr})
combined with the sum rule in (\ref{Fsr}). In Fig.~\ref{fig:phifuns},
we show the results for the functions $\phi_i(w)$ obtained by varying
the parameters $\omega_c$ and $T$ in the ranges described above. For
all three functions, we observe a mild decrease with $w$. We note
that these results can be trusted only for moderate values of $w$,
which is however sufficient for all practical purposes. For $w\gg 1$,
the non-perturbative contributions to the sum rules would rapidly
vanish once the non-locality of the condensates was taken into
account.

\section{Concluding Remarks}

We have presented a simultaneous determination of the HQET parameters
$\lambda_1$ and $\lambda_2$ from a QCD sum-rule analysis of a 3-point
correlation function containing the local operator $\bar
h_{v'}\Gamma\,i g_s G^{\mu\nu} h_v$ together with two heavy-light
currents. Our sum rule for the parameter $\lambda_2$, which describes
the chromo-magnetic interaction of a heavy quark, agrees with a
result derived previously in Ref.~\cite{BaBr}. However, our sum rule
for $\lambda_1$ is new. It incorporates the virial theorem, which
relates the kinetic energy of a heavy quark inside a meson to its
chromo-electric interaction with gluons \cite{virial}. We shall argue
below that our approach is superior to that of Ball and Braun
\cite{BaBr}, who extracted $\lambda_1$ from the correlator of the
kinetic operator $\bar h_v\,(i D_\perp)^2 h_v$ with two heavy-light
currents. The reason is that the virial theorem makes explicit an
``intrinsic smallness'' of $\lambda_1$, which is otherwise hidden
by a large background from excited-state contributions. Our numerical
results are given in (\ref{lam2mu0}) and (\ref{lam1val}). The value
of $\lambda_1$ implies an average residual momentum of the heavy
quark inside a meson of order 200--400~MeV; the result for
$\lambda_2$ translates into a spin splitting of
$m_{B^*}^2-m_B^2=(0.60\pm 0.12)~\mbox{GeV}^2$, which is in good
agreement with experiment.

We like to add a final comment regarding the difference between our
approach and that of Ball and Braun \cite{BaBr}, who obtained a much
larger value for $\lambda_1$ than our result in (\ref{lam1val}). The
reason is that their sum rule contains a large contribution from a
``bare'' quark loop, which is $O(\alpha_s^0)$. Before the continuum
subtraction, their result reads\footnote{For simplicity, we do not
display the terms of $O(\alpha_s)$ here, since the main problem is to
understand the origin of the leading term.}
\begin{equation}
   - \lambda_1\,F^2\,e^{-2\bar\Lambda/T} + C(T)
   = \frac{9 T^5}{4\pi^2}
   - \frac 38\,m_0^2\langle\bar q q\rangle + O(\alpha_s) \,,
\label{BBsr}
\end{equation}
where $C(T)$ represents the contributions of excited states, which
are removed in the conti\-nu\-um subtraction. We have argued in
Ref.~\cite{virial} that the virial theorem, which relates $\lambda_1$
to a matrix element of the gluon field-strength tensor, does not
allow terms not containing the gauge coupling in the sum rule for
$\lambda_1$. We shall now explain how this statement is consistent
with (\ref{BBsr}).

To start with, let us stress that we do not claim that the authors of
Ref.~\cite{BaBr} made a calculational mistake; indeed, we have
checked that their result (\ref{BBsr}) is correct. What we are going
to argue is that the leading perturbative term on the right-hand side
of (\ref{BBsr}), together with part of the contribution from the
mixed condensate, must not be attributed to the ground-state, but
rather to excited states. The virial theorem helps to avoid from the
start any subtleties related to the complicated problem of the
continuum subtraction in 3-point sum rules.

In order to explain our argument, it is necessary to go into some of
the details of QCD sum-rule calculations in the HQET. The sum rules
for the matrix elements of local dimension-5 operators are closely
related to the derivatives of the sum rules for some
lower-dimensional operators with respect to the Borel parameters.
Consider the 3-point sum rules for the meson decay constant (i.e.\
the sum rule for the Isgur--Wise function evaluated at zero recoil)
and for the product $\xi_-(1) F^2$, where $\xi_-(1)$ is the
zero-recoil limit of a form factor defined in terms of the matrix
element of the operator $\bar h_{v'}\Gamma iD^\mu h_v$ \cite{Luke}.
These sum rules read \cite{IWfun}
\begin{eqnarray}
   F^2\,e^{-2\bar\Lambda(t+t')} + C_1(t+t')
   &=& \frac{3}{4\pi^2}\,\frac{1}{(t+t')^3} - \langle\bar q q\rangle
    + \frac{m_0^2\langle\bar q q\rangle}{4}\,(t+t')^2 + O(\alpha_s)
    \,, \nonumber\\
   \xi_-(1)\,F^2\,e^{-2\bar\Lambda(t+t')} + C_2(t,t')
   &=& \frac{3}{16\pi^2}\,\frac{7 t'-t}{(t+t')^5}
    - \frac{m_0^2\langle\bar q q\rangle}{24}\,(7 t'-t)
    + O(\alpha_s) \,,
\label{xixi}
\end{eqnarray}
where $t=1/\tau$ and $t'=1/\tau'$ are the two Borel parameters
associated with the variables $\omega$ and $\omega'$, and $C_i$
denote the contributions to the correlators from exited states. As a
consequence of the orthogonality of states, $C_1(t+t')$ is a function
of the sum of the Borel parameters only. Note, however, that the
second
sum rule is not symmetric in the two Borel variables. This is a
consequence of the fact that the operator whose matrix element
defines $\xi_-(w)$ contains a derivative acting on one of the
heavy-quark fields. Consequently, the function $C_2(t,t')$ is not
symmetric in its arguments. Taking derivatives with respect to the
Borel parameters $t$ and $t'$, we can derive a set of related sum
rules containing powers of the parameter $\bar\Lambda$. In the case
of the first sum rule in (\ref{xixi}), it clearly does not matter
whether we take a derivative with respect to $t$ or $t'$; however,
the same statement is not true in the case the second sum rule. After
taking the derivatives, we set the the Borel parameters equal,
$t=t'=1/2T$, in which case the first sum rule in (\ref{xixi}) reduces
to (\ref{Fsr}). We obtain
\begin{eqnarray}
   F^2\,e^{-2\bar\Lambda/T} + C_1(1/T)
   &=& \frac{3 T^3}{4\pi^2} - \langle\bar q q\rangle
    + \frac{m_0^2\langle\bar q q\rangle}{4 T^2} + O(\alpha_s) \,,
    \nonumber\\
   \bar\Lambda\,F^2\,e^{-2\bar\Lambda/T} - \frac 12\,C_1'(1/T)
   &=& \frac{9 T^4}{8\pi^2}
    - \frac{m_0^2\langle\bar q q\rangle}{4 T} + O(\alpha_s) \,,
    \nonumber\\
   \bar\Lambda^2\,F^2\,e^{-2\bar\Lambda/T} + \frac 14\,
   C_1''(1/T) &=& \frac{9 T^5}{4\pi^2}
    + \frac{m_0^2\langle\bar q q\rangle}{8} + O(\alpha_s) \,,
\label{xisums}
\end{eqnarray}
and
\begin{eqnarray}
   \xi_-(1)\,F^2\,e^{-2\bar\Lambda/T} + C_2(1/2T,1/2T)
   &=& \frac{9 T^4}{16\pi^2}
    - \frac{m_0^2\langle\bar q q\rangle}{8 T} + O(\alpha_s) \,,
    \nonumber\\
   \bar\Lambda\xi_-(1)\,F^2\,e^{-2\bar\Lambda/T} - \frac 12\,
   \partial_t C_2(1/2T,1/2T) &=& \frac{3 T^5}{2\pi^2}
    - \frac{m_0^2\langle\bar q q\rangle}{48} + O(\alpha_s) \,,
    \nonumber\\
   \bar\Lambda\xi_-(1)\,F^2\,e^{-2\bar\Lambda/T} - \frac 12\,
   \partial_{t'} C_2(1/2T,1/2T) &=& \frac{3 T^5}{4\pi^2}
    + \frac{7}{48}\,m_0^2\langle\bar q q\rangle + O(\alpha_s) \,.
\label{ximisums}
\end{eqnarray}
It is crucial that the continuum contribution $C_2(t,t')$ is not
symmetric in $t$ and $t'$; otherwise the last two sum rules would be
inconsistent with each other.

Using the equations of motion of the HQET, one can show that
$\xi_-(1)=\bar\Lambda/2$ \cite{Luke}. In fact, this relation follows
by comparing the first sum rule in (\ref{ximisums}) with the second
sum rule in (\ref{xisums}), provided we identify $C_2(1/2T,1/2T) =
-\frac 14\,C_1'(1/T)$. This relation between the continuum
contributions is indeed satisfied when one adopts the standard
continuum model, according to which
\begin{equation}
   C_1(1/T) = \frac{3 T^3}{4\pi^2}\,\left[ 1 - \delta_2(\omega_c/T)
   \right] \,, \quad
   C_2(1/2T,1/2T) = \frac{9 T^4}{16\pi^2}\,\left[ 1
   - \delta_3(\omega_c/T) \right] \,.
\end{equation}
Things are more subtle for the last two sum rules in
(\ref{ximisums}), which have the same ground-state contribution.
Inserting the normalization condition $\xi_-(1)=\bar\Lambda/2$, we
find that their sum agrees with the last sum rule in (\ref{xisums}).
However, the only logical explanation for the fact that the
theoretical expressions on the right-hand sides of these sum rules do
not coincide is that the difference between these expressions
contributes to the exited states only, but not to the ground state.
Hence, taking the difference between the two sum rules in
(\ref{ximisums}) leads to a sum rule with vanishing ground-state
contribution. It reads
\begin{equation}
   (\partial_{t'} - \partial_t)\,C_2(1/2T,1/2T)
   = \frac{3 T^5}{2\pi^2} - \frac{m_0^2\langle\bar q q\rangle}{3}
   + O(\alpha_s) \,.
\end{equation}

Let us now come back to the sum rule (\ref{BBsr}) for the matrix
element of the kinetic operator derived by Ball and Braun. We can
combine it with the above relation in such a way that the
contribution of the bare quark loop, which is forbidden by the virial
theorem, is eliminated from the right-hand side of the sum rule. Then
the result takes the form
\begin{equation}
   - \lambda_1\,F^2\,e^{-2\bar\Lambda/T} + \tilde C(T)
   = \frac{m_0^2\langle\bar q q\rangle}{8} + O(\alpha_s) \,,
\end{equation}
where the new continuum contribution is given by $\tilde C(T)=C(T) -
\frac 32(\partial_{t'} - \partial_t)\,C_2(1/2T,1/2T)$. What we have
achieved is to identify the leading perturbative contribution in
(\ref{BBsr}), as well as part of the contribution of the mixed
condensate, as a contribution to the exited states coupling to the
correlation function. What remains is nothing but our sum rule for
$\lambda_1$ obtained in (\ref{lamsr}), with the correct coefficient
in front of the mixed condensate.

\vspace{0.3cm}
{\it Acknowledgements:\/}
It is a pleasure to thank Zoltan Ligeti and Chris Sachrajda for
helpful discussions.

\end{document}